\documentclass[11pt]{article}
\usepackage{apalike,latexsym,amssymb,bbm}

\def\mytitle{A Generalized Equivalence Principle}

\def\ben{\begin{enumerate}}
\def\een{\end{enumerate}}
\def\bi{\begin{itemize}}
\def\ei{\end{itemize}}
\def\bq{\begin{quote}}
\def\eq{\end{quote}}
\def\be{\begin{equation}}
\def\ee{\end{equation}}

\def\setE{\mathbbm E}
\def\setP{\mathbbm P}
\def\setR{\mathbbm R}
\def\Mink{\setR^{(1,3)}}              
\def\DiffM{{\cal D\mbox{\em iff}(M)}} 

\renewcommand{\baselinestretch}{1.1}
\begin{document}
\sloppy

\thispagestyle{empty}

\vspace*{-22mm}

\noindent {\em International Journal of Modern Physics D}, Vol. 9,
No. 6 (2000) 633-647\\
(E-print arXiv:gr-qc/0004054, v2 with only a few corrections).

\vspace{12mm}

\noindent{\Large\bf \mytitle}

\vspace{7mm}

\hfill\parbox{13cm}{

{\sc Holger Lyre}

\vspace{4mm}

{\small
Institut f\"ur Philosophie,
Ruhr-Universit\"at Bochum,
D-44780 Bochum,
Germany,\\
e-mail: holger.lyre@ruhr-uni-bochum.de
}

\vspace{7mm}

April 2000

\vspace{3mm}

{\small
\paragraph{Abstract.}
Gauge field theories may quite generally be defined as describing
the coupling of a matter-field to an interaction-field, and they are
suitably represented in the mathematical framework of fiber bundles.
Their underlying principle is the so-called {\em gauge principle},
which is based on the idea of deriving the coupling structure
of the fields by satisfying a postulate of local gauge covariance.
The gauge principle is generally considered to be sufficient
to define the full structure of gauge-field theories.
This paper contains a critique of this usual point of view:
firstly, by emphazising an {\em intrinsic gauge theoretic conventionalism}
which crucially restricts the conceptual role of the gauge principle and,
secondly, by introducing a new {\em generalized equivalence principle}
-- the identity of {\em inertial and field charge}
(as generalizations of inertial and gravitational mass)
-- in order to give a conceptual justification for combining
the equations of motion of the matter-fields
and the field equations of the interaction-fields.
}

}\hspace*{5mm}

\vspace{5mm}

{\small
\bq
\renewcommand{\contentsname}{Content}
\tableofcontents
\eq
}

\newpage


\section{The gauge principle}

At least three of the four known fundamental interactions are
representable as gauge-field theories (for short, `gauge theories').
Thus, from the conceptual point of view it is of particular interest
to understand the role, effect and explanatory power
of the possible underlying principles of this kind of theories.
Now, it is quite customary to consider the so-called
{\em gauge principle} (GP) as the one and only and, hence,
sufficient principle to almost dictate
the full structure of gauge-field theories
by merely satisfying a postulate of local gauge covariance.
However, as we shall see, this turns out too strong a statement.

It is the purpose of the first three sections of this paper
to present a detailed analysis of the GP's true conceptual role.
To begin with, recall the {\sc Dirac-Maxwell} theory as the paradigm
gauge theory. Note that it is the basic idea of gauge-field
theories to represent the combination of two field theories:
a matter-field and an interaction-field or, in our example,
the {\sc Dirac} and the {\sc Maxwell} fields.\footnote{Accordingly,
quantum electrodynamics is the quantized version of
the {\sc Dirac-Maxwell} gauge theory. Note, however, that it is
misleading to call {\sc Maxwell}'s theory, i.e. classical electrodynamics,
a gauge theory merely because of its gauge freedom
$A'_\mu(x) = A_\mu(x) - \partial_\mu \Lambda(x)$;
since a true gauge theory couples matter-fields and
interaction-fields. I shall strictly use this terminology;
cf. {\sc Lyre} (1999).}
In the spirit of the GP we must start with a free matter-field theory,
represented by the {\sc Dirac} Lagrangian
\be
\label{dirac}
{\cal L}_D = \bar\psi(x) \left(i \gamma^\mu \partial_\mu - m \right) \psi(x) ,
\ee
where each spinor component $\psi^i$ is a $U(1)$ representation.
This Lagrangian is clearly invariant under global $U(1)$ gauge
transformations $\psi'(x) = e^{i q^{(i)} \Lambda } \psi(x)$
(the factor $q^{(i)}$ and its notation will become clear below).
For the construction of a $U(1)$ gauge theory we demand
covariance\footnote{Throughout this paper we speak
of {\em covariance of theories} as opposed to {\em invariance of
certain objects of a theory} (with respect to some symmetry group).}
of the free matter-field theory represented by (\ref{dirac})
under local (i.e. spacetime-dependent) transformations
\be
\label{psilocal}
\psi(x) \ \to \ \psi'(x) = e^{i q^{(i)} \Lambda(x) } \psi(x)
\equiv \hat U (x) \ \psi(x) .
\ee
This requirement is called the {\em gauge postulate}.
Inserting (\ref{psilocal}) in (\ref{dirac}) we get an extraneous term
$- q^{(i)} \bar\psi ( \gamma^\mu \partial_\mu \Lambda ) \psi$,
which may be compensated by introducing a vector field
\be
\label{replacepot}
A_\mu(x) = \partial_\mu \Lambda(x)
\ee
with the transformation behaviour
\be
\label{potlocal}
A'_\mu(x) = A_\mu(x) - \partial_\mu \Lambda(x) .
\ee
In other words, in order to satisfy the postulate of local gauge
covariance of the free theory, we have to replace the usual derivative
by the so-called {\em covariant derivative}
\be
\label{covderiv}
\partial_{\mu} \ \to \ D_{\mu} = \partial_{\mu} + i q^{(i)} \ A_\mu .
\ee
Thus, instead of (\ref{dirac}), we get a transformed {\sc Dirac} Lagrangian
\be
\label{dirac'}
{\cal L}'_D = \bar\psi(x) \left(i \gamma^\mu D_\mu - m \right) \psi(x) ,
\ee
which is invariant under the combined
local gauge transformations (\ref{psilocal}) and (\ref{potlocal}).

Mathematically, gauge theories allow for an adaequate representation
within the geometrical framework of principal fiber bundles
$\setP$ and their associated vector bundles $\setE$.
The bundle approch defines a typical correspondence between
the geometrical and the physical terms arising in gauge theories:
Gauge potentials are connections of $\setP$, local sections
of $\setE$ represent matter-fields. The gauge group is the structure
group of the bundle and its generators are the gauge bosons.
Local gauge transformations may be regarded as automorphisms of $\setP$.
Finally, the bundle curvature is to be considered as the
interaction-field strength.


\section{The intrinsic gauge theoretic conventionalism}

Usually, the replacement (\ref{replacepot}) is interpreted
as the introduction of a new ``interaction-field''.
Hence, the transformed Lagrangian (\ref{dirac'}) is considered to
describe a theory in which the {\sc Dirac} matter-field is coupled
to a ``field'' $A_\mu$. Given $A_\mu$, it is possible to form
the bundle curvature tensor
\be
\label{curvature}
F_{\mu\nu} = \partial_\mu A_\nu - \partial_\nu A_\mu .
\ee
On the background of the interaction interpretation we are motivated
to call $A_\mu$ the {\em gauge potential} and $F_{\mu\nu}$,
its derivative, the {\em gauge-field strength}.\footnote{In theoretical
high energy physics it is common practice to call $A_\mu$ a ``field'',
but in order to avoid confusion I shall consistently call $A_\mu$
the gauge potential (actually, $F_{\mu\nu}$ describes the real
gauge-fields $\vec E$ and $\vec B$).}
This interpretation is suggested by the following consideration.
As mentioned in the previous section, the gauge postulate basically
amounts to replacing global by local transformations.
Due to {\sc Noether}'s first theorem global gauge invariance of
some Lagrangian under a $k$-dimensional {\sc Lie} group
is connected with the existence of $k$ conserved currents.
For the one-dimensional $U(1)$ we obtain the electric current
\be
\label{j-i}
\jmath^{(i)}_\mu = q^{(i)} \ \bar\psi \gamma_\mu \psi
\ee
in a notation analogous to $q^{(i)}$. It obeys the continuity equation
\be
\label{conti}
\partial_\mu \jmath^{(i) \, \mu} = 0 .
\ee
Hence, because of the existence of a charged {\sc Noether} current
it is indeed suggestive to assume that there exists a true interaction
field which couples to this current and which may be represented by
its potential $A_\mu$. More explicitly,
\be
\label{L_int-i}
{\cal L}^{(i)}_{int} = - \jmath^{(i)} _\mu A^\mu
\ee
should be considered as the coupling part of the transformed
Lagrangian ${\cal L}'_D$, now written as
\be
\label{dirac+int}
{\cal L}'_D = {\cal L}_D + {\cal L}^{(i)}_{int}
\ee
to underline the interaction interpretation.

Note that all of the above derivations arise from the GP
``with a little help'' from {\sc Noether}'s theorem.
From this point of view, the GP appears as a most powerful tool -- at
first glance sufficient to derive the coupling to an interaction-field!
But, did it really do this job?
Let us reconsider the crucial step of the argument:
the local gauge transformation (\ref{psilocal}) implies an extraneous
term basicially consisting of the derivation of the phase function
$\Lambda(x)$. However, the ``replacement'' (\ref{replacepot})
is of course nothing but a mere renaming.
It is nice heuristics to assume $A_\mu$ as a real potential
which couples to a charged current $j^{(i)}_\mu$
-- but we are, at this stage, by no means forced to do so.
All that has happend is that the Lagrangian was transformed
in a way that it looks as if it implied a coupling.
What the GP rather leads to is a special sort of an internal
coordinate transformation in the local fibers of the bundle space.
The coupling may very well be understood as a mathematical
artefact arising from a mere bundle coordinate transformation
and, hence, representing a pure conventional manoeuvre!
I propose to call this an {\em intrinsic gauge theoretic conventionalism}.
It has no physical significance so far.

To obtain a theory with true physical potentials we need
to secure a non-vanishing field strength, which -- as a dynamical
element and thereby truly changing the physical situation --
must obey some field equations.\footnote{One might, however,
argue that the celebrated {\sc Aharonov-Bohm} effect
as well as other topological effects in gauge theories point to
the physical significance of the potentials alone.
This may be conceded, but it does not concern our issue here.
We want to understand the {\em principles} underlying the
{\em full} structure of gauge-field theories.
The AB effect may then be described as a special result.
Moreover, although the effect seems to show a direct
influence of the gauge potential on the electron matter-field
(in a setting with vanishing field strength),
the occurrence of this potential outside the solenoid is only due
to the fact that there is a real interaction-field inside the solenoid.
Without the existence of this field, of course, there is no AB effect
at all. It is only the existence of the internal magnetic field
that justifies the interpretation of the vector potential
external to the solenoid as a true physical interaction potential.}
To underline this point one should recall the bundle structure
of gauge theories.
Quite generally, the types of bundles arising in gauge theories
may be distinguished in a three-fold way:
\bi
\item type 1: Trivial bundles with flat connections,
\item type 2: Trivial bundles with non-flat connections,
\item type 3: Non-trivial bundles.
\ei
This distinction is a useful one from the point of view
of characterizing the possible bundle geometries in gauge theories.
Fiber bundles can be considered as generalizations of the direct product
of spaces -- here: base space and typical fiber.
Only locally do they look like a product space. Trivial bundles,
however, can even globally be reduced to a direct product.
Therefore, type 1 bundles really turn out as superfluous structures.
Of course, with mathematical rigour the same is true
for type 2 bundles, too. Only type 3 bundles appear to be justified
(from this point of view it is not necessary to distinguish between
non-trivial bundles with flat or non-flat connections).\footnote{From
the conceptual point of view, however, even type 2 bundles
(which actually mainly occur in the experimentally vindicated
part of the standard model)
may be considered as extremely useful, since gauge theoretic bundles
in general allow for a natural distinction between the merely
mathematical structures, which live in the base space,
and the truly physical structures of gauge-field theories,
which live in the fibers; cf. {\sc Guttmann} and {\sc Lyre} (2000).}

Now, a flat connection is defined as a connection with vanishing
curvature and therefore physically describing a field-free
configuration (i.e. vanishing gauge-field strength).
What kind of bundle geometry is then described
by the transformed Lagrangian ${\cal L}'_D$ in (\ref{dirac'})
or (\ref{dirac+int}), respectively?
As we saw there is no definite answer to this question,
since on the basis of ${\cal L}'_D$ alone we simply cannot decide
whether we are dealing with a truly coupled or merely internally
transformed Lagrangian. This, again, means that the GP by itself
just implies a pure conventional coordinate transformation
(in a trivial bundle) -- a procedure which does not distinguish
between type 1 and type 2 bundles.
Thus, in order to give rise to type 2 (or maybe even type 3) bundles
we have to introduce a real interaction-field dynamics.
This shall be analyzed in the next section.


\section{Gauge field equations}
\label{gfe}

As pointed out in the previous section it is tempting to consider
$A_\mu$ and $F_{\mu\nu}$ already as a true gauge potential
and a true gauge-field strength.
This temptation becomes even higher, since we indeed already
know a field which is describable by a tensor of type (\ref{curvature})
and a corresponding potential with gauge freedom (\ref{replacepot}).
In this respect there also exists a true field charge,
which we denote by $q^{(f)}$.
This theory, of course, is {\sc Maxwell}'s electrodynamics.

As a propagating field, the electromagnetic field $F_{\mu\nu}$
obeys {\sc Maxwell}'s vacuum equations which come in two pairs.
The first pair can be derived from (\ref{curvature})
as a mere {\sc Bianchi} identity
\be
\label{bianchi}
\partial_{[\mu} F_{\nu\rho]} = 0 .
\ee
The second pair equations are given by
\be
\label{vacuum}
\partial_\mu F^{\mu\nu} = 0
\ee
and turn out as the {\sc Euler-Lagrange} equations of
\be
\label{maxwell_L}
{\cal L}_M = - \frac{1}{4} F_{\mu\nu} F^{\mu\nu} .
\ee

Indeed, the {\sc Maxwell} Lagrangian ${\cal L}_M$ is gauge invariant
under (\ref{replacepot}) and most textbooks insist
on the statement that (\ref{maxwell_L}) is the simplest Lagrangian
with this property. Moreover, under the usual field theoretic requirements
such as limitation to second order field equations, for instance,
(\ref{maxwell_L}) may even turn out uniquely.
But still, from the GP alone we are not {\em forced}
to consider the bundle curvature $F_{\mu\nu}$ as a
true electromagnetic field with its own kinematics (\ref{vacuum}).
Rather, to describe the field {\em dynamics} we have to introduce
field sources. In {\sc Maxwell}'s theory these are the electric charges.
Written in {\sc Lorentz} covariant notation the field charges
are represented by a four current density
\be
\label{j-f}
\jmath^{(f)}_\mu = \rho^{(f)} v_\mu
\ee
with $\rho^{(f)}$ denoting the density of the field charge $q^{(f)}$
per volume element $V$ and $v_\mu$ as the four velocity of $V$.
We get a dynamical coupling of the form
\be
\label{L_int-f}
{\cal L}^{(f)}_{int} = - \jmath^{(f)}_\mu A^\mu .
\ee
Thus, the {\sc Euler-Lagrange} equations for the dynamical Lagrangian
of electrodynamics ${\cal L}_{ED} = {\cal L}_M + {\cal L}^{(f)}_{int}$
become
\be
\label{inhom-maxwell}
\partial^\mu F_{\mu\nu} = \jmath^{(f)}_\nu .
\ee

To be sure, Lagrangians (\ref{L_int-i}) and (\ref{L_int-f}),
which describe the coupling, have indeed the same structure,
and it is {\em very} tempting to make the identification
\be
\label{L-int-equiv}
{\cal L}^{(i)}_{int} = {\cal L}^{(f)}_{int} \equiv {\cal L}_{int} .
\ee
This would allow us to combine the vacuum field Lagrangian
${\cal L}_M$ with the coupled matter-field Lagrangian ${\cal L}'_D$
to get the total Lagrangian of the {\sc Dirac-Maxwell} theory
\be
\label{DMtheory}
{\cal L}_{DM} = {\cal L}_D + {\cal L}_{int} + {\cal L}_M .
\ee
In fact, ${\cal L}_{DM}$ describes new physics compared to
${\cal L}_D$ or ${\cal L}'_D$, respectively.
The entire {\sc Dirac-Maxwell} theory represents the properly
combined dynamics of matter-field and gauge-field -- the key feature
of gauge-field theories. Nonetheless, nothing in the argument structure
of the GP {\em forces} us to make the crucial identification
(\ref{L-int-equiv}) based on the assumption of the {\em equivalence}
of $\jmath^{(i)}_\nu$ and $\jmath^{(f)}_\nu$.
Admittedly, the GP works nicely as a heuristic guiding principle
to introduce the field Lagrangian and its dynamical coupling.
But we are not allowed to sweep the quite different origin
of currents (\ref{j-i}) and (\ref{j-f}) under the rug,
for this is far too strong a physical assumption,
which does not flow out of a principle of merely conventional
character -- such as the GP.

Thus, in order to obtain the full {\sc Dirac-Maxwell} theory
(\ref{DMtheory}) we need a truly {\em physical} principle.
Otherwise, there remains a ``missing link'' -- at least from
the foundational point of view.
In search for such a linking principle it might be useful
to take a lesson from general relativity, first.


\section{The lesson from general relativity}

\subsection{The gauge theoretic analogy}

General relativity theory (GR) is the commonly accepted theory
of gravitation. As it stands it does not have a completely consistent
gauge-field theory structure, but rather shows the most important gauge
theoretic features such as a principal fiber bundle geometry\footnote{In
standard GR the principal bundle is the bundle of orthonormal frames
(tetrad bundle), the associated vector bundle is the tangent bundle
of the pseudo-{\sc{Riemann}}ian spacetime manifold \cite{trautman80a}.}
and -- crucial for our considerations -- an underlying gravitational GP.
The most distinctive feature of gravitation compared to other gauge
field theories is, besides its classical rather than quantum nature,
the so-called {\em soldering} of the bundle base space and its fibers:
in gravitational theories the bundle geometry is inseparably connected
to the geometry of the base space. In other words, gravitation is the field
which represents the metric properties of spacetime itself
(rather than of some internal symmetry space).
For our purposes, however, it is sufficient to recall the gauge
theoretic analogy of GR.\footnote{For a philosophical account on
gravitational gauge theories see {\sc Eynck} and {\sc Lyre} (2001);
the ``ultimate'' physical reference is {\sc Hehl} et al. (1995).}

Let us try to mimic a suitable gravitational GP.
We start with flat {\sc Minkowski} space $\Mink$,
i.e. the interaction-free case in which no gravitation exists.
Hence, we consider the free geodesic equation
\be
\label{free_geodesic}
m^{(i)} \ \frac{d v^\mu(\tau)}{d \tau} = 0 .
\ee
The four vector $v^\mu(\tau) = \frac{d x^\mu(\tau)}{d \tau}$,
tangent to a timelike curve $x^\mu(\tau)$, denotes the velocity
of a (pointlike) particle with inertial mass $m^{(i)}$.
Note that $v^\mu \equiv \theta_0^\mu$ together with a system
of three spacelike vectors $\theta_i^\mu$ forms a tetrad
$\theta^\mu_\alpha=(\theta^\mu_0, \theta^\mu_1, \theta^\mu_2, \theta^\mu_3)$.
A tetradial reference frame may be considered as representing
an observer in spacetime. Clearly, (\ref{free_geodesic}) is
globally invariant under the homogeneous {\sc Lorentz} group $SO(1,3)$
(in principle, other groups are possible as well,
such as the group of {\sc Poincar\'e} translations $\Mink$,
the diffeomorphism group $\DiffM$ etc.).
Now, the local gauge postulate amounts to\footnote{We use
three types of indices:
\renewcommand{\baselinestretch}{0.5}
\normalsize\footnotesize
\ben
\item curved (holonomic) {\sc Lorentz} indices $\mu, \nu \ldots$
in the space-time manifold (external space),
\item flat (anholonomic) tetrad indices $\alpha, \beta \ldots$
in local {\sc Minkowski} space (internal space), and
\item the Latin index $a$ in the {\sc Lie} algebra of the gauge group.
\een
\renewcommand{\baselinestretch}{1.1}
\normalsize\footnotesize
}
\be
\label{GT2a}
\theta^\mu_\alpha(\tau)
= \hat L_\alpha^\beta(x) \ \theta^\mu_\beta(\tau)
= e^{ (\hat M^a)_\alpha^\beta \Lambda^a(x) }
\ \theta^\mu_\beta(\tau)
\ee
with the 6 generators $\hat M^a$ of $SO(1,3)$
and corresponding local, i.e. spacetime-dependent,
transformation parameters $\Lambda^a(x)$
and $x \equiv x^\mu(\tau)$. In the geodesic equation of
(\ref{GT2a}) we substitute
\be
  (\hat M^a)_\alpha^\beta \ \frac{d}{d \tau} \Lambda^a(x)
= (\hat M^a)_\alpha^\beta \ \partial_\nu \Lambda^a(x) \ \frac{d}{d \tau} x^\nu (\tau)
= \Gamma^\beta_{\nu\alpha}(x) \ \frac{dx^\nu (\tau)}{d \tau} .
\ee
$\Gamma_\nu$ must itself satisfy local $SO(1,3)$ gauge transformations
\be
\label{GT2b}
\Gamma^\gamma_{\nu\alpha}(x)
= \hat L^\gamma_\alpha(x) \ \Gamma^\delta_{\nu\gamma}(x) \ (\hat L^{-1})_\delta^\gamma(x)
- \hat L^\delta_\alpha(x) \ \partial_\nu                 \ (\hat L^{-1})_\delta^\gamma(x) .
\ee
We insert (\ref{GT2a}) and (\ref{GT2b}) in (\ref{free_geodesic})
to get the geodesic equation in curved space-time
\be
\label{curv_geodesic}
m^{(i)} \ \frac{d}{d \tau} \theta^\mu_\alpha(\tau)
+ m^{(i)} \ \Gamma^\beta_{\nu\alpha}
\frac{d x^\nu(\tau)}{d \tau} \theta_\beta^\mu(\tau) = 0 .
\ee
In other words, the gauge postulate concerning local {\sc Lorentz} rotations
of the tetrads leads to a covariant derivative $\nabla_\tau$, such that
\be
\frac{d}{d \tau} \theta^\mu_\alpha(\tau) \
\to \ \nabla_\tau \theta^\mu_\alpha(\tau)
= \frac{d}{d \tau} \theta^\mu_\alpha(\tau) + \Gamma^\beta_{\nu\alpha}
\frac{d x^\nu(\tau)}{d \tau} \theta_\beta^\mu(\tau) .
\ee

It is now straightforward to compare GR with the {\sc Dirac-Maxwell}
gauge theory as discussed above.
Obviously, in the gravitational case the gauge potentials are given
by the {\sc Levi-Civita} connection $\Gamma_\mu$
(with {\sc Christoffel} symbols as its components).
From the gravitational potential we may form the curvature tensor
$R^\alpha_{\mu\nu\beta}$, the so-called {\sc Riemann} tensor,
which represents the gravitational field strength.
Is it, then, justified to interpret (\ref{curv_geodesic}) as the
geodesic equation in {\em curved} spacetime -- and, thus, to interpret
$\Gamma_\mu$ as a true gravitational potential which gives rise to
a true gravitational field strength?
Of course, the answer can be no other than negative!
Actually, in GR it is even easier to realize the purely conventional
character of the connection occuring in the GP:
it is already very well-known that non-vanishing {\sc Christoffel}symbols
may occur simply because of some peculiar choice of coordinates
(i.e. curvilinear coordinates in flat space).
Of course, this does not mean that we already have non-vanishing
curvature and, thus, a true gravitational field.
Things are more obvious in GR than in gauge theories in general
since the intrinsic gauge theoretic conventionalism applies to
{\em external} rather than to internal coordinate transformations
(again, a peculiarity of the bundle soldering).
Thus, from the mere logic of the GP we still remain
in a type 1 case: flat connections with vanishing fields.
It is therefore necessarry to obtain new physics by appealing to
a real physical principle which GR is based upon:
the principle of equivalence of inertia and gravitation.


\subsection{The equivalence principle}

In {\sc Newton}ian physics mass appears in three different meanings,
firstly, as a source of the gravitational potential, the so-called
{\em active mass} $m^{(a)}$ in
\be
\label{m_a}
\Phi_g = - \frac{m^{(a)}}{r} ,
\ee
secondly, as a quantity on which gravitation acts, the so-called
{\em passive mass} $m^{(p)}$ in
\be
\label{m_p}
\vec F = - m^{(p)} \vec\nabla \Phi_g ,
\ee
and, thirdly, as the tendency of resistance of motion, the
so-called {\em inertial mass} $m^{(i)}$ in
\be
\vec F = m^{(i)} \ddot{\vec r} .
\ee
Due to {\em ``actio=reactio''} the equality of active and passive mass,
identified as the so-called {\em gravitational mass} $m^{(g)}$,
is concluded
\be
m^{(a)} = m^{(p)} \equiv m^{(g)}.
\ee
The equality of gravitational and inertial mass, however,
does not follow from the {\sc Newton}ian axioms.
In other words, the {\em empirical fact} of the mass-independence
of the free fall of ponderable matter remains unexplained
\be
\label{free-fall}
\ddot{\vec r} = - \underbrace{\frac{m^{(g)}}{m^{(i)}}}_{= 1} \vec\nabla \Phi_g .
\ee
As is well-known, {\sc Einstein} took this empirical fact very serious:
an observer in a freely falling elevator does not feel his own weight.
Also, feeling weight is not sufficient to tell him whether this is
due to an acceleration of the elevator or due to the existence of
an external (homogeneous) gravitational field.
Thus, inertia and gravitation, {\sc Einstein} concluded, are
{\em identical in essence} (in German: ``wesensgleich'').
The equivalence principle (EP) acquires the following form:
{\em There always exists a local frame of reference,
such that the gravitational field vanishes.}

I should note two points:
firstly, the above formulation is stronger than the empirical universality
of the free fall. The latter merely requires that in a gravitational
field all (point) particles follow the same path irrespective of their
mass. This statement is sometimes called the {\em weak EP}.
As opposed to this, {\sc Einstein}'s EP is a {\em strong EP}.
It includes the weak version, but, moreover, states that gravity has
no influence on any physical process whatsoever.
In the following, we will use this version (simply refering to it as EP).
Now, secondly, the above formulation contains special relativity as
a limiting case. As we will see shortly, this becomes a cruicial
feature for generalizing the EP to gauge theories.

But another step first:
We are now prepared to return to our question at the end
of the preceeding subsection.
It is indeed the EP which allows for an interpretation of the
{\sc Levi-Civita} connection $\Gamma_\mu$ as a true
gravitational potential {\em in the general case}.
By analogy with section \ref{gfe} we must consider the gravitational field
with its own field dynamics given by the {\sc Einstein} field equations
\be
\label{einstein}
R_{\mu\nu} - \frac{1}{2} \, R \ g_{\mu\nu} = - \kappa \ T_{\mu\nu} .
\ee
Analogous to the inhomogeneous {\sc Maxwell} equations
(\ref{inhom-maxwell}), the l.h.s. of (\ref{einstein})
is made of curvature (and its contractions, respectively).\footnote{To
be sure, the analogy is incomplete. The GR Lagrangian
${\cal L}_{GR} = \frac{1}{2 \kappa} \sqrt{-g} R + {\cal L}_{Matter}$
is linear and not quadratic in the fields, compared to (\ref{maxwell_L}).
It is possible, however, to have a full analogy to the general
gauge theoretic {\sc Yang-Mills} approach (cf. section \ref{ymcc})
in the framework of an alternative gravitational gauge theory of
the full {\sc Poincar\'e} group with a quadratic field Lagrangian
including curvature and torsion \cite{hehl_etal80}. Standard GR,
however, does neither consistently fit into this picture,
nor does the equation of motion (\ref{curv_geodesic}) arise
from a genuine field theoretic Lagrangian density with an
action functional ${\cal S} = - m^{(i)} \int d x^4 {\cal L}$
(such as (\ref{dirac}), for instance), but rather
${\cal S} = - m^{(i)} \int d \tau \sqrt{g_{\mu\mu} v^\mu v^\nu}$;
see also {\sc Eynck} and {\sc Lyre} (2001).}
The r.h.s. of both equations represents the field source.
Thus, the gravitational mass $m^{(g)}$, which is encoded in the
energy-momentum tensor $T_{\mu\nu}$, acts as the field charge
of gravity. Due to the equivalence of inertial and gravitational mass,
\be
\label{equiv-mi-mg}
m^{(i)} = m^{(g)} ,
\ee
we are now able to combine the geodesic equation
of motion (\ref{curv_geodesic}), which includes $m^{(i)}$,
with the {\sc Einstein} field equations (\ref{einstein}),
which include $m^{(g)}$.
From the conceptual point of view, the identification
(\ref{equiv-mi-mg}) -- a direct result of the EP --
is the one and decisive link between the equation of motion,
which originates from the GP, and the field equations,
which are merely motivated by the GP in their formal structure.
Due to the EP the {\sc Levi-Civita} connection
may be identified {\em for the general case}
with a non-flat connection representing a true gravitational potential
with non-vanishing fields, i.e. curvature.
In other words, only the EP allows for type 2 bundles
to enter gravitational theories.


\section{A generalized equivalence principle}

\subsection{The ``missing link''}

From all of the above it seems very natural to ask whether there exists
a principle in gauge theories in general, which, like the EP in
gravitational theories, links up equations of motion and field equations.
And if so, maybe it can even be formulated in such a way that the
gravitational EP turns out as a special case for gravitational gauge
theories. It is the final purpose of this paper to demonstrate exactly
this statement: the idea of a {\em generalized equivalence principle}.

Recall the prerelativistic analogy of {\sc Newton}ian gravitation
and electrodynamics.
If we simply replace the gravitational potential $\Phi_g$
by the {\sc Coulomb} potential $\Phi_e$,
we may introduce an active electric charge, $\Phi_e = - \frac{q^{(a)}}{r}$,
analogous to (\ref{m_a})
as well as a passive one, $\vec F = -q^{(p)} \vec\nabla \Phi_e$,
analogous to (\ref{m_p})
and, again due to {\em actio=reactio}, we identify
\be
q^{(a)} = q^{(p)} \equiv q^{(f)} ,
\ee
where $q^{(f)}$ is the field charge occuring in section \ref{gfe}.
Quite generally, $q^{(f)}$ denotes the field source of the gauge-fields
and, hence, gravitational mass may be looked upon as the special case
for gravitational interaction.

We must now ask for a generalization of $m^{(i)}$.
Is it conceivable to have something like an {\em ``inertial charge''}?
In quantum gauge-field theories the equations of motion originate
from the {\sc Dirac} Lagrangian (\ref{dirac}).
This is the reason why the $q^{(i)}$-notation was chosen already in
equations (\ref{psilocal}), (\ref{covderiv}) and (\ref{j-i}).
Curiously, the inertial charge implicitly occurs as a factor
in the phase of the matter wavefunction $\psi e^{i q^{(i)} \Lambda }$.
Due to the intrinsic gauge theoretic conventionalism this factor
remains purely conventional without any physical significance
up to the point when we plug in the crucial identification
\be
\label{equiv-qi-qf}
q^{(i)} = q^{(f)} .
\ee
This statement, in analogy to (\ref{equiv-mi-mg}), originates from
what I propose to call a generalized EP. Here is a formulation
equivalent to the strong version of the gravitational EP:
\bq
Generalized equivalence principle (GEP):
{\em It is always possible to perform a local gauge transfomation
such that, locally (i.e. at a point), the gauge-field vanishes.}
\eq
Thus, any gauge-field theory includes the interaction-free theory
as a local limiting case.
The above formulation of the GEP implies that the local gauge
transformations are now considered as irrevocably connected
with the occurrence of an interaction field, which has its origin
in the field charges and obeys its own dynamics.
In other words: As far as solely the GP is concerned,
local gauge transformations must be considered as bundle
coordinate transfromations. Hence, the GP merely amounts
to a passive interpretation of local gauge transfomations.
As soon as we take the GEP, a truly physical principle,
under consideration, too, we are forced to interpret
local gauge transfomations actively.\footnote{In more detail,
{\sc Michael Redhead} (1998) has discussed certain problems arising
in connection with active and passive interpretations
of local gauge transformations.
It seems very likely that our presentation sheds new light
on his discussion in favour of an active interpretation of
gauge transformations. This will be the task of a special paper.}


\subsection{Yang-Mills theories, charges and couplings}
\label{ymcc}

So far, I have solely discussed the {\sc Dirac-Maxwell} theory
(or QED, respectively) as well as the GR analogue as gauge theories.
However, the quantum gauge theories of the standard model are
non-abelian gauge theories, known as {\sc Yang-Mills} theories.
I shall show now that the above considerations
apply to {\sc Yang-Mills} theories in general.
The non-abelian groups occuring in the standard model are
$SU(2)_F$ for flavour charge (weak isospin) and
$SU(3)_C$ for color.
In the general $SU(n)$-case, the gauge postulate leads to
local gauge transformations
\be
\Psi(x) \ \to \ \Psi'(x) = e^{i g^{(i)} \Lambda^a(x) \hat t^a} \Psi(x)
\equiv \hat U (x) \ \Psi(x)
\ee
instead of (\ref{psilocal}),
where $\Psi=(...,\psi^i,...)^T$, $i=1, 2 ... n$ denotes
the fundamental spinor representation of $SU(n)$
with $a=1,2...n^2-1$ generators $\hat t^a$.
As a generalization of (\ref{potlocal}) we get
\be
\label{YM-potlocal}
B^{\prime a}_\mu(x) \ \hat t^a
= \hat U(x) \ B^a_\mu(x) \ \hat t^a \ \hat U^+(x)
 - \frac{i}{g^{(i)}} \ \hat U(x) \ \partial_\mu \ \hat U^+(x) .
\ee
The covariant derivative reads
\be
D_{\mu} = \partial_{\mu} + i g^{(i)} \ B^a_\mu \ \hat t^a .
\ee
We may, again, construct a field strength tensor analogous to
(\ref{curvature})
\be
\label{YMcurvature}
F^a_{\mu \nu} \ \hat t^a = - \frac{i}{g^{(i)}} \ [ D_\mu , D_\nu ]
= \Big( \partial_\mu B^a_\nu - \partial_\nu B^a_\mu
  - g^{(i)} f^{abc} B^b_\mu B^c_\nu \Big) \ \hat t^a .
\ee
From global gauge covariance under $SU(n)$ we get $n^2-1$
conserved {\sc Noether} currents
\be
\label{YMnoether}
\jmath_\mu^{(i) \, a} = g^{(i)} \ \bar\Psi \gamma_\mu \hat t^a \Psi
\ee
with a continuity equation
\be
(D^\mu \jmath_\mu^{(i)})^a
= \partial^\mu \jmath_\mu^{(i) \, a}
- g^{(i)} f^{abc} \ B^{\nu b} \ j_\nu^{(i) \, c} = 0 .
\ee
Comparing this equation with its abelian counterpart (\ref{conti})
we can read off the decisive difference between abelian and
non-abelian gauge theories: in the latter not only the matter-field
current is conserved, there rather exists a contribution from
the gauge potentials $B^a_\mu$ which carry charge.
Also, the third term in (\ref{YMcurvature}) represents the
self-interaction of the {\sc Yang-Mills} gauge-bosons.

In analogy to the QED case we have introduced an inertial charge
$g^{(i)}$ (or rather a coupling constant, see remarks below).
But we still do not have any field equations, since the curvature
tensor (\ref{YMcurvature}) is not necessarly non-zero from
the GP's point of view. Instead, we must introduce the so-called
{\sc Yang-Mills} equations, generalizations of
the {\sc Maxwell} equations (\ref{inhom-maxwell}),
\be
\label{inhom-YM}
(D^\mu F_{\mu\nu})^a
= \partial^\mu F^a_{\mu\nu} - g^{(f)} f^{abc} \ B^{\mu b} \ F^c_{\mu\nu}
= \jmath_\nu^{(f) \, a}.
\ee
There is also a {\sc Bianchi} identity analogous to (\ref{bianchi}),
and a {\sc Yang-Mills} Lagrangian ${\cal L}_{YM}$
analogous to (\ref{maxwell_L}).
In (\ref{inhom-maxwell}), there arises a field charge $g^{(f)}$.
To be sure, from the point of view of pure {\sc Yang-Mills} field
theory formulae (\ref{YM-potlocal}) and (\ref{YMcurvature}),
which also arise there, should then be written with $g^{(f)}$,
either. The identification
\be
g^{(i)} = g^{(f)}
\ee
expresses the GEP. Only on this basis are we allowed
to combine both theories entering a true {\sc Dirac-Yang-Mills}
gauge theory ${\cal L}_{DYM} = {\cal L}'_D + {\cal L}_{YM}$.

It may be useful to recall the different meanings
of charges and coupling constants in field physics.
In QED, for instance, the electric charge $q$ is measured
in units $e$, the elementary charge.\footnote{More advanced,
one uses dimensionless coupling strengths, such as
$\alpha \sim \frac{1}{137}$ for QED or
$\alpha_s \sim \frac{1}{10}$ for QCD.
Of course, in physics only the dimensionless ratios of
the fundamental parameters such as masses and charges
will have any convention-independent significance.}
At the same time, $e$ plays the role of the
electromagnetic coupling constant. In {\sc Yang-Mills} theories,
however, we discover a so-called {\em universality of the coupling}.
This includes two points:
firstly, the coupling of the $B^a_\mu$ gauge potential to the
matter-field current (\ref{YMnoether}) has the same strength
as the $B^a_\mu$-$B^a_\mu$ coupling, i.e. the coupling of the charged
gauge bosons among each other.
Secondly, any particle, which couples weakly or strongly, couples
with the same interaction strength.
This means that there are no multiples of weak or color charges.
Interestingly, the first point appears now as a consequence of the GEP
rather than the GP (although, the GP would be impossible otherwise).
The second point does not follow from any of both principles.
For in QED there are ``multiples'', more precisely thirds, of $e$:
the electromagnetic coupling is $-\frac{1}{3}e$ for d-s-b-quarks,
$\frac{2}{3}e$ for u-c-t-quarks, and $1e$ for electrons, muons or taus.

An explanation of this latter point may rather be due to a proper
understanding of the quantum nature of the fundamental charges.
With respect to this, consider gravitational theories.
It seems that we may clearly distinguish between mass as the
charge of gravity and the coupling constant $\kappa= 8 \pi G$.
But a simple reason for this could be the fact that we are still
dealing with a classical theory, i.e. we have not quantized mass.
If so, we perhaps would be able to introduce an analogue of $e$
(or $\frac{1}{3}e$, respectively) such as the {\sc Planck} mass,
for instance.
But these questions are certainly beyond the scope of this paper.


\subsection{Experimental conclusions}

By its very nature the equivalence principle must be checked
empirically, since it is the one and decisive empirical input
to gauge theories.
Indeed, the identification (\ref{equiv-qi-qf}) turns out
as a consequence of the GEP. The reason is that if we regard
the connection field as truly physical, it must have its
sources in the field charges. But this exactly means that
we have to identify the $q^{(i)}$ factor arising in the
transformation (\ref{psilocal}) with the field charge $q^{(f)}$.
Now, if we assume, for a moment, the ratio of $q^{(f)}$ and $q^{(i)}$
to deviate from $1$, then this means that different types of particles
of equal electric charge would couple differently to the
electromagnetic field. Hence, we should expect a difference in the
coupling of electrons and muons, d-quarks and s-quarks etc.
But this is certainly not what we observe!
Therefore, the GEP predicts a whole variety of {\em null-experiments}.

Note that the analogy to the gravitational case is quite
straightforward. We must check empirically that different
materials do have the same free fall behaviour.
Otherwise, the $m^{(g)}$-$m^{(i)}$-ratio in (\ref{free-fall})
would lead to different results for, say, 1 kilogram of wood
than for 1 kilogram of iron.
In the same sense as this possibility, from {\em a priori} grounds,
cannot be excluded, the coupling of different types of particles
with the same gauge field charge, on principle, cannot be excluded.
In other words, the GEP is equivalent to the statement
that all types of particles with the same $q^{(f)}$ show the same
coupling.
Otherwise, we would have to write down a different {\sc Dirac} equation
\be
\left(i \gamma^\mu \partial_\mu - m \right) \psi e^{i q^{(i)} \Lambda}
= c_p \ q^{(f)} \ \gamma^\mu A_\mu \ \psi e^{i q^{(i)} \Lambda}
\ee
for different types of particles with the same $q^{(f)}$,
but with a particle type-dependent factor $c_p$.\footnote{Of course,
the situation is more confusing in the quantum gauge theoretic case
due to the occurence of $q^{(i)}$ in the phase of the
matter-wavefunction $\psi e^{i q^{(i)} \Lambda }$.
Conceptually I decidedly agree with what {\sc Sunny Auyang} (1995)
has pointed out: in modern field theories the very notion
of an event becomes inseparably connected with the idea
of an interaction.
The inertial charge seems to be no direct observable (in contrast
to inertial mass). However, for the formulation of the GEP it
is sufficient to consider the ratio of $q^{(i)}$ and $q^{(f)}$.}
High energy physics tells us that this is, of course, not the case;
and therefore we may safely say that the GEP is empirically very well
confirmed.


\section{Concluding remarks}

It was the purpose of this paper to convince the reader that
the celebrated GP cannot take the burden of a full derivation
of gauge field theories.
Rather, the crucial gauge-theoretic combination
of a pure matter-field and a pure gauge-field theory,
i.e. ${\cal L}_{total} = {\cal L}'_D + {\cal L}_{GF}$,
should be based on a second principle,
a generalized equivalence principle.
This is a new idea, which of course does not change
the practical application of gauge theories,
but rather affects their deeper conceptual understanding.

The idea of postulating an EP as a general principle in gauge theories
has already been expressed by {\sc Gerhard Mack} (1981).
However, as far as I can see, his arguments have a quite different
origin: {\sc Mack} seems to be motivated by the close analogy to GR.
Certainly, he did not stress the intrinsic gauge theoretic
conventionalism. I do not see, however, that there is anything
in his presentation which must convince the reader to apply his EP
beyond the GP, if one does not make the conventionalism claim.

From the standpoint of philosophy of physics,
{\sc Harvey Brown} (1999) and {\sc Paul Teller} (2000a)
have emphasized that the gauge argument
applies to a mere coordinate transformation. Thus, in a way
they also take the gauge theoretic conventionalism serious.
Moreover, {\sc Teller} (2000b) pointed out that from the mere GP
the QED connection appears to be only the $A_\mu$ vector field
rather than $q A_\mu$.
He then concludes that the GP does not lead to a derivation of the
charge $q$ (or coupling constant $e$, respectively).
I do not follow the whole variety of suggestions which he draws from
his analysis, but certainly he comes very close to my point.
From this paper's perspective, of course, one could argue the
following way: there is a factor, say $q^{(i)}$, which is not
determined from the gauge argument.
On the other hand, there is a {\sc Maxwell} field source $q^{(f)}$.
But we will know about their identification from experiment only.

Hence, the GEP stresses the one and decisive empirical input
for considering gauge theories as physical theories.


\newpage

\section*{Acknowledgements}

Special thanks to
{\sc M. Drieschner}, {\sc T.~O. Eynck}, {\sc D. Graudenz},
{\sc M. Kuhlmann} and {\sc M. Tielke}
for helpful discussions and several valuable comments
on earlier versions of the manuscript.

\vspace*{2mm}


\bibliographystyle{apalike}


\end{document}